\documentstyle[11pt]{article}
  
\newcommand{\ds}{\displaystyle }
\newcommand{\R}{{\sf R\hspace*{-0.9ex}%
\rule{0.15ex}{1.5ex}\hspace*{0.9ex}}}
\newcommand{\Z}{{\sf Z\hspace*{-0.9ex}%
\rule{0.15ex}{1.5ex}\hspace*{0.9ex}}}
\newcommand{\N}{{\sf N\hspace*{-0.9ex}%
\rule{0.15ex}{1.5ex}\hspace*{0.9ex}}}

\title{A self-similar field theory for 1D linear elastic continua and self-similar diffusion problem}

\author{ {\sl Thomas M.  Michelitsch$^{1}$\footnote{Corresponding author, e-mail~: michel@lmm.jussieu.fr }, G\'{e}rard A. Maugin$^{1}$}\\ \\ {\sl Mujibur Rahman$^2$, Shahram Derogar$^3$} \\ \\ {\sl Andrzej F. Nowakowski$^4$, Franck C. G. A. Nicolleau$^4$} \\ \\
$^1$ Universit\'{e} Pierre et Marie Curie, Paris 6\\
Institut Jean le Rond d'Alembert \\ CNRS UMR 7190 \\ France
 \\ \\ $^2$General Electric Company\\
Aviation \\ USA
 \\ \\
$^3$ School of Mechanical, Aerospace and Civil Engineering\\
The University of Manchester
\\
United Kingdom
\\ \\
$^4$ Sheffield Fluid Mechanics Group, Department of Mechanical Engineering\\
University of Sheffield\\
United Kingdom
\\ \\  \\ {\it Submitted to Journal of Differential Equations} \\ \\ {\small \it \jobname .tex }
}
\begin{document}

\maketitle

\paragraph{Abstract}
This paper is devoted to the analysis of some fundamental problems of linear elasticity
in 1D continua with self-similar interparticle interactions. We introduce a self-similar
continuous field approach where the self-similarity is reflected by equations of motion which are spatially non-local convolutions with power-function kernels (fractional integrals).
We obtain closed-form expressions for the static displacement Green's function due to a unit $\delta$-force. 
In the dynamic framework we derive the solution of the {\it Cauchy problem} and the retarded Green's function. We deduce the distribution of a self-similar variant of diffusion problem with L\'evi-stable distributions as solutions with infinite mean fluctuations describing the statistics  L\'evi-flights.
We deduce a hierarchy of solutions for the self-similar Poisson's equation which we call "self-similar potentials".
These non-local singular potentials are in a sense self-similar analogues to the 1D-Dirac's $\delta$-function.
The approach can be the starting point to tackle a variety of scale invariant interdisciplinary problems.


\section{INTRODUCTION}

The discovery of fractal geometry in nature by Beno\^it Mandelbrot in the seventies of the last century had a great impact on science and opened us a new view on nature \cite{mandel,mandel1,mandel2,sapoval}.
In the last decade also an increasing interest has risen in the mechanics of materials
having a microstructure which is exactly or statistically invariant over a wide range of scales. Such materials
can be in a good approximation idealized as "fractal" materials. Such an idealization
is to be understood in the same way as the notion of the "infinite medium". Despite there is no infinite medium in nature, this notion has proven to be a highly appropriate idealization.
Some interesting results to describe the mechanical properties in fractal materials have been achieved recently \cite{kigami,bodarenko,ostoja,epstein}. However a generally accepted theory of "fractal mechanics" is not yet established so far. Therefore simple even simplistic models capturing only a few aspects of the characteristic fractal behavior in materials are highly desirable.

The point of departure in the present paper is the continuum limit of the 1D medium which has been defined as "quasi-continuous linear chain" in \cite{michel}. Before we start with our demonstration we should like to specify "our" notion of "self-similarity": The
notion self-similarity used in the present paper as well as in \cite{michel} corresponds to the notion of "{\it self-similarity at a point}" commonly used in the mathematical literature \cite{peitgen}. An object is self-similar in the strict sense if it {\it consists} of parts which are exact rescaled copies of the entire object. In contrast is the notion "self-similarity at a point" (with "point\footnote{Where that "point" is a fixed-point of that scaling operation.} we mean a space-point which is contained by or attached to the object, where we place for example the origin of a coordinate system): An object which is "self-similar at a point" {\it contains only a single part} which is a re-scaled copy of the entire object \cite{peitgen}.
The notion of "self-similarity at a point" implies that the scaling invariance repeats over an infinity of scales. In this paper as well as in \cite{michel} we call an object just "self-similar" without always mentioning that we actually mean "self-similar at a point". To be more precise: When we call a function $\Lambda(h)$ self-similar with respect to $h$ which we define by $\Lambda(Nh)=N^{\delta}\Lambda(h)$ for a prescribed scaling factor $N$, then strictly speaking the function
$\Lambda(h)$ is self-similar at the point $h=0$. The simplest self-similar functions of this type are power-functions $h^{\delta}$.

The paper is organized as follows: In section \ref{deltaforce}
we deduce {\it static} Green's function of displacements due to a $\delta$-unit force in closed form.
In section \ref{dynamique} we construct the basic solution of the Cauchy problem in the form of two integral kernels which give the solutions due to prescribed initial displacement- and velocity-field, respectively, in analogy to the classical d'Alembert's solution. One of these two kernels
determine also the retarded time-domain dynamic Green's function.

In section \ref{transport} we analyze a self-similar diffusion problem which we formulate
as an initial value problem for a given initial distribution by utilizing our self-similar Laplacian. The solutions of this problem turn out to be L\'evi-stable distributions with infinite mean fluctuations. As a spin-off result we obtain a hierarchy of functions which we call "self-similar potentials`` as they play the analogue role for our self-similar Laplacian as the Newtonian potentials do in the case of a "traditional" Poisson-equation. These hierarchy of potentials includes also the {\it static Green's function} as a solution of a $\delta$-type source.

\section{The self-similar elastic continuum}
\label{1D}

We consider a purely elastic medium in 1D with a spatial continuous constant mass distribution
of density $1$ where any spatial point $x$ represents a material point which interacts harmonically with an ensemble of other material points located at $x\pm hN^s$ ($s\in \Z_0$)
in such a way that the elastic energy-density at $x$ is a self-similar function with respect to $h$ (at point $h=0$). This is expressed below by relation (\ref{selfsimh0}). The fractal properties of the dispersion relation of such a system have been considered in our recent paper \cite{michel}.
As a point of departure we invoke some results from that paper: \newline
The Hamiltonian of that material system is

\begin{equation}
\label{Hamiltonian}
H=
\frac{1}{2}\int_{-\infty}^{\infty}\left(\dot{u}^2(x,t)+{\cal V}(x,t,h)\right){\rm d}x
\end{equation}
where $x$ denotes the space- and $t$ the time coordinates.
${\cal V}(x,t,h)$ indicates the elastic energy density\footnote{The additional factor ${1}/{2}$ in the elastic
energy avoids double counting.}

\begin{equation}
\label{elastic}
{\cal V}(x,t,h)=\frac{1}{2}\sum_{s=-\infty}^{\infty}N^{-\delta s}\left[(u(x,t)-u(x+hN^s,t))^2
+(u(x,t)-u(x-hN^s,t))^2\right]
\end{equation}
which converges in the range $0<\delta<2$ and where we assume $h>0$ and $N$ being a prescribed scaling factor. Without loss of generality we can restrict ourselves to $N>1$ ($N\in \R$).
In (\ref{Hamiltonian}), (\ref{elastic}) $u$ and $\dot{u}=\frac{\partial }{\partial t}u$ stand for the displacement field and the velocity field, respectively. (\ref{elastic}) has the property of being self-similar with respect to $h$ at point $h=0$, namely

\begin{equation}
\label{selfsimh0}
{\cal V}(x,t,Nh)=N^{\delta}{\cal V}(x,t,h)
\end{equation}

As a point of departure for the approach to be developed we evoke the continuum limit of (\ref{elastic})
and the resulting equation of motion. We define the continuum limit as $N=1+\zeta$ ($0<\zeta<<1$) so $\tau=hN^s$ becomes a continuous variable and we can
write a self-similar function $\Lambda(h)$ which fulfills a self-similarity condition (\ref{selfsimh0}) asymptotically as \cite{michel}

\begin{equation}
\label{selfu}
\Lambda(h)=\sum_{s=-\infty}^{\infty}N^{-\delta s}f(N^sh) \approx \frac{h^\delta}{\zeta}\int_0^{\infty}\frac{f(\tau)}{\tau^{\delta +1}}{\rm d}\tau
\end{equation}
having the form of a power function $\Lambda(h)=const\,h^{\delta}$.
Both the discrete as well as the continuous representation of (\ref{selfu}) converge for sufficiently good functions (see details in \cite{michel}).
From (\ref{selfu}) follows that in that continuum limit we can write the elastic energy density (\ref{elastic}) as a functional of the displacement field $u(x,t)$ in the form

\begin{equation}
\label{elastfunc}
{\cal V}(x,t,h)=\frac{h^\delta}{2\zeta}\int_0^{\infty}\frac{\left\{(u(x,t)-u(x+\tau,t))^2+
(u(x,t)-u(x-\tau,t))^2\right\}}
{\tau^{\delta +1}}{\rm d}\tau
\end{equation}
which exists as in the discrete case (\ref{elastic}) in the band $0<\delta<2$. The equation of motion (self-similar wave equation) has then the form \cite{michel}

\begin{equation}
\label{fractlapself} \frac{\partial^2 }{\partial t^2}u(x,t) = \Delta_{(\delta,h,\zeta)}u(x,t)
\end{equation}

The right hand side of (\ref{fractlapself}) can be conceived as the {\it self-similar Laplacian} of the medium. The self-similar Laplacian is necessarily a non-local and {\it self-adjoint} negative definite operator and is obtained as

\begin{equation}
\label{laplace}
\Delta_{(\delta,h,\zeta)}u(x)=\frac{h^{\delta}}{\zeta}\int_0^{\infty}\frac{(u(x-\tau)+u(x+\tau)-2u(x))}{\tau^{1+\delta}}\,{\rm
d}\tau \,,\hspace{2cm} 0<\delta<2
\end{equation}
 and exists in the interval $0<\delta<2$. It might be sometimes convenient to rewrite (\ref{laplace}) in the equivalent form
\begin{equation}
\label{laplaceb}
\Delta_{(\delta,h,\zeta)}u(x)=\frac{h^{\delta}}{\zeta\delta}\frac{d }{d x}\int_0^{\infty}\frac{(u(x+\tau)-u(x-\tau))}{\tau^{\delta}}\,{\rm
d}\tau \,,\hspace{2cm} 0<\delta<2
\end{equation}

We can express Laplacian (\ref{laplace}), (\ref{laplaceb}) in terms of {\it Weyl-Marchaud fractional
derivatives} which are defined as \cite{weyl-marchaud}
\begin{equation}
\label{weylmleft}
D_l^{\delta}u(x)=\frac{\delta}{\Gamma(1-\delta)}\int_0^{\infty}\frac{(u(x)-u(x-\tau)}{\tau^{\delta+1}}
\,{\rm d}\tau \,,\hspace{1cm}{\rm left-sided} \,,\hspace{1cm} 0<\delta<1
\end{equation}
and
\begin{equation}
\label{weylmright}
D_r^{\delta}u(x)=\frac{(-1)^{\delta}\delta}{\Gamma(1-\delta)}\int_0^{\infty}\frac{(u(x)-u(x+\tau)}{\tau^{\delta+1}}
\,{\rm d}\tau \,,\hspace{1cm}{\rm right-sided} \,,\hspace{1cm} 0<\delta<1
\end{equation}
where $0<\delta<1$ indicates the range of existence of each fractional integral $D_{l,r}^{\delta}u(x)$, respectively. $\Gamma(z)$ indicates the $\Gamma$-function (faculty-function) \cite{abramowitz}

\begin{equation}
\label{gammafu}
\Gamma(z+1)=:z!=\int_0^{\infty}e^{-\tau}\tau^z{\rm d}\tau ,\hspace{2cm} Re(z) > -1
\end{equation}
The condition $Re(z)>-1$ is required that integral (\ref{gammafu}) exists.
Then the Laplacian (\ref{laplace}) has the representation\footnote{We have $-1=e^{i\pi}$ with
$(-1)^{-\delta}=e^{-\pi i \delta}$ as we use throughout this paper for any complex number $z=|z|e^{i\varphi}$ the {\it principal value} $-\pi < \varphi=Arg(z) \leq \pi$ for its argument $\varphi$.}

\begin{equation}
\label{laplaceweylmarchaud}
\Delta_{(\delta,h,\zeta)}u(x)=-\frac{h^{\delta}\Gamma(1-\delta)}{\zeta\delta}\left(D_l^{\delta}+ (-1)^{-\delta}D_r^{\delta}\right)u(x)\,,\hspace{2cm} 0<\delta<1
\end{equation}
where the actual range of existence of (\ref{laplaceweylmarchaud}) is $0<\delta<2$ when we write both terms as one single integral like in (\ref{laplace}).
We should like to mention that the representation (\ref{laplaceweylmarchaud}) in terms of standard fractional integrals is not crucially helpful in order to solve problems. Therefore, we should rather conceive Laplacian (\ref{laplace}), (\ref{laplaceb}) itself as definition of a self-adjoint combination of fractional derivatives.

The goal of this paper is to analyze some static and dynamic problems which are governed by the self-similar Laplacian (\ref{laplace}), (\ref{laplaceb}).
The following section is devoted to deduce an explicit expression for the static displacement Green's function.

\section{Static Green's function}
\label{deltaforce}

This paragraph is devoted to deduce the {\it static} Green's function for the displacement field due to a $\delta$-point force. We define this Green's function $g(x)$ by

\begin{equation}
\label{staticgreen}
\Delta_{(\delta,h,\zeta)}g(x)+\delta(x)=0
\end{equation}
where $\delta(x)$ denotes Dirac's $\delta$-function.
The static displacement field $u(x)$ due to a force density $f(x)$ is then defined by
\begin{equation}
\label{defeqdispl1}
\Delta_{(\delta,h,\zeta)}u(x)+f(x)=0
\end{equation}
and can be represented as a convolution
\begin{equation}
\label{Gfcon1}
u(x)=\int_{-\infty}^{\infty}g(x-\tau)f(\tau){\rm d}\tau
\end{equation}
with the static Green's function $g(x)$ of (\ref{staticgreen}) as convolution-kernel.
By introducing the Fourier transforms

\begin{equation}
\label{fuourierdisplace}
g(x)=\frac{1}{2\pi}\int_{-\infty}^{\infty}{\tilde g}(k)e^{ikx}{\rm d}k
\end{equation}
and
\begin{equation}
\label{deltafunc}
\delta(x)=\frac{1}{2\pi}\int_{-\infty}^{\infty}e^{ikx}{\rm d}k
\end{equation}
it follows from (\ref{staticgreen}) that
\begin{equation}
\label{gfoutrafo}
{\tilde g}(k)=\frac{1}{\omega^2(k)}
\end{equation}

By taking into account that $e^{ikx}$ are the eigenfunctions of the $\Delta_{(\delta,h,\zeta)}$-operator (\ref{laplace}) with $\Delta_{(\delta,h,\zeta)}e^{ikx}=-\omega^2(k)e^{ikx}$ the dispersion relation is obtained from 

\begin{equation}
 \label{omega2}
\omega^2(k)= -\frac{h^{\delta}}{\zeta}\int_0^{\infty}\left(e^{ik\tau}+e^{-ik\tau}-2\right)\tau^{-\delta-1}{\rm d}\tau
\end{equation}
which was also given in \cite{michel} in the form
\begin{equation}
\label{integralauxiliare}
A_{\delta}=\frac{2h^{\delta}}{\zeta}\int_0^{\infty}\frac{(1-\cos(s))}{s^{1+\delta}}{\rm d}s \,,\hspace{2cm} 0<\delta<2
\end{equation}

This integral exists in the same interval $0<\delta<2$ just as the Laplacian (\ref{laplace}).
We use now the identity\footnote{Re(..) denotes the real- and Im(..) the imaginary part of a complex number (..).}

\begin{equation}
 \label{idenfac} |\tau|^{-\delta-1}=\frac{1}{\cos{\frac{\pi(\delta +1)}{2}}}Re\{\lim_{\epsilon\rightarrow 0+}(\epsilon-i\tau)^{-\delta-1}\}
=\lim_{\epsilon\rightarrow 0+}\frac{1}{\delta !\cos{\frac{\pi(\delta +1)}{2}}}Re\{\int_0^{\infty}e^{-s(\epsilon -i\tau)}s^{\delta}{\rm d}s\}
\end{equation}
which is well defined in the range $0<\delta<2$ of existence of the Laplacian.
Replacing identity (\ref{idenfac}) in the integral (\ref{omega2}) yields\footnote{See details on integrals of this type
in appendix \ref{balphamin1}.}

\begin{equation}
 \label{omega3}
\omega^2(k)= -\frac{h^{\delta}\pi}{\zeta\delta ! \cos{\frac{\pi(\delta +1)}{2}}}\int_0^{\infty}s^{\delta}\left\{\delta(s-|k|)-2\delta(s)\right\}{\rm d}s=\frac{h^{\delta}\pi}{\zeta\delta ! \sin{\frac{\pi\delta}{2}}}|k|^{\delta}\,,\hspace{2cm} 0<\delta<2
\end{equation}
which has the form \cite{michel}

\begin{equation}
\label{dispersionrel}
\omega^2(k)= A_{\delta} |k|^{\delta} \,,\hspace{2cm} 0<\delta<2
\end{equation}
where coefficient $A_{\delta}$ is given by 

\begin{equation}
\label{Adelta2}
A_{\delta}=\frac{h^{\delta}}{\zeta}\frac{\pi}{\delta ! \sin{(\frac{\pi\delta}{2})}} >0 \,,\hspace{2cm} 0<\delta < 2
\end{equation}
We conclude that
the positiveness of the constant $A_{\delta} >0$ (which is determined by $\sin{\frac{\pi\delta}{2}} >0$) of (\ref{Adelta2}) due to the positiveness of integral (\ref{integralauxiliare}) reflects nothing but the condition of elastic stability, i.e. $\omega^2(k)>0$ for $k\neq 0$ which is a {\it physically necessary} condition. This condition together with the condition of existence of Laplacian (\ref{laplace}) constitute a {\it physically sufficient condition} which gives the $\delta$-range of {\it physical consistency} namely the range
$0<\delta<2$. We emphasize that the regularization performed in (\ref{idenfac}) would allow to obtain converging integrals $\forall \delta > -1$. However among these {\it mathematically admissible} $\delta$-values only those within $0<\delta < 2$ lead to {\it physically meaningful results}. This is true for all {\it physical} problems formulated with Laplacian (\ref{laplace}).
\newline\newline
\noindent{\bf Explicit Green's function}:\newline
The integral (\ref{fuourierdisplace}) assumes with (\ref{gfoutrafo}) and (\ref{dispersionrel})
the form

\begin{equation}
\label{fuourierdisplace2}
g(x)=\frac{1}{2\pi A_{\delta}}\int_{-\infty}^{\infty}|k|^{-\delta}e^{ikx}{\rm d}k
\end{equation}
where we make use of the relation

\begin{equation}
 \label{kmidelta}
|k|^{-\delta}=\frac{1}{(\delta-1)!\cos{\frac{\delta\pi}{2}}}\lim_{\epsilon\rightarrow 0 +}Re\{\int_0^{\infty}e^{-\tau(\epsilon-k)}\tau^{\delta-1}{\rm d}\tau\}
\end{equation}
which is well defined in the range of existence $0<\delta< 2$ of the Laplacian (\ref{laplace}).
Using this identity to evaluate (\ref{fuourierdisplace2}) we finally arrive
at
\begin{equation}
\label{reform3}
g(x)= g_0\,|x|^{\delta -1}
\end{equation}
with the pre-factor

\begin{equation}
\label{constant1}
g_0=\frac{1}{2A_{\delta}(\delta-1)!\cos{\frac{\pi\delta}{2}}}=\frac{\zeta \delta }{2\pi h^{\delta}}\frac{\sin{\frac{\delta\pi}{2}}}{\cos{\frac{\delta\pi}{2}}}\,,\hspace{2cm} 0<\delta<2
\end{equation}

Expression (\ref{reform3}) holds for $0<\delta<2$ which yields physical consistent behavior for the deformation $\frac{d}{dx}g(x)$ having the exponent $\delta-2 <0$: $|\frac{d}{dx}g(x\rightarrow 0)|\rightarrow \infty$ being singular at $x=0$ and $|\frac{d}{dx}g(x\rightarrow \pm \infty)|\rightarrow 0$ vanishing for $|x|\rightarrow\infty$. The consistence of expressions (\ref{reform3}), (\ref{constant1}) with (\ref{gfoutrafo}) can be directly verified
by performing the back transformation

\begin{equation}
\label{backtrafo}
{\tilde g}(k)=\int_{-\infty}^{\infty}g(x)e^{-ikx}\,{\rm d}x= 2g_0\lim_{\epsilon\rightarrow 0+}Re\{\int_0^{\infty}|x|^{\delta -1}e^{-x(\epsilon-ik)}\,{\rm d}x\}
\end{equation}
which can be rewritten as

\begin{equation}
\label{rewriteexpress}
{\tilde g}(k)= 2g_0\Gamma(\delta)\cos{\frac{\pi \delta}{2}}|k|^{-\delta}
\end{equation}
This expression must coincide with the expression ${\tilde g}(k)=\frac{1}{\omega^2(k)}$ of (\ref{gfoutrafo}) with the dispersion relation (\ref{dispersionrel}). This leads to
the condition that

\begin{equation}
\label{condnec}
\frac{1}{A_{\delta}}=2g_0\Gamma(\delta)\cos{\frac{\pi \delta}{2}}
\end{equation}
which is fulfilled by (\ref{constant1}). It is possible to evaluate (\ref{fuourierdisplace2}) in the range $0<\delta<1$
directly to arrive at 
\begin{equation}
 \label{g02} g(x)= \frac{1}{\pi A_{\delta}}\lim_{\epsilon\rightarrow 0+}Re\{\int_0^{\infty}e^{-k(\epsilon-ix)}|k|^{-\delta}{\rm d}k\}=\frac{(-\delta) !}{\pi A_{\delta}}Re(\epsilon-ix)^{\delta-1}=\frac{\Gamma(1-\delta)\sin{\frac{\pi\delta}{2}}}{\pi A_{\delta}}|x|^{\delta-1}
\end{equation}

Comparision of $g_0$ of (\ref{g02}) and (\ref{constant1}) lead to the condition that

\begin{equation}
\label{eulerrelation}
\Gamma(\delta)\Gamma(1-\delta)=\frac{\pi}{\sin{\pi \delta}}
\end{equation}
which is known as {\it Euler's reflection formula} \cite{abramowitz}. Consequently the correctness of expressions (\ref{reform3}), (\ref{constant1}) for the Green's function and (\ref{gfoutrafo}) is proven. 
For the analysis to follow it is useful to generalize the definition of the $\Gamma$-function extending it to $\alpha < -1$
in such a way that for the extended definition of $\alpha !$ Euler-relation (\ref{eulerrelation}) is fulfilled\footnote{This definition of $\alpha !$ maintains the representations of regularized integrals of the form (\ref{balpahalph}) for all admissible values $\alpha \in \R$.} 
\begin{equation}
 \label{expresgen}
\alpha! = \Gamma(\alpha+1)=-\frac{\pi}{\Gamma(-\alpha)\sin{\pi\alpha}} 
\end{equation}
where $\Gamma(-\alpha)$ is well-defined by (\ref{gammafu}) for $\alpha<-1$. Definition (\ref{expresgen}) is motivated in appendix \ref{balphamin1} for $\alpha < -1$.
Equation (\ref{expresgen}) complements the usual definition (\ref{gammafu}) of the $\Gamma$-function by extending it to the ``forbidden'' $\alpha$-range of (\ref{gammafu}) $\alpha < -1$.

We again empasize our previous observation that the Green's function (\ref{reform3})ff. is {\it physically meaningful} only in the range $0<\delta<2$ to obtain
consistent behavior for the {\it deformation} $\frac{d}{dx}g(x)$ behaving as $|x|^{\delta-2}$, such as its vanishing for $|x|\rightarrow \infty$ and its singular behavior for $x\rightarrow 0$.

\section{Cauchy problem and dynamic Green's functions}
\label{dynamique}

In this section we construct some fundamental solutions of dynamic problems: First in paragraph \ref{cauchy} we construct the solution of the {Cauchy problem} and in paragraph \ref{dyngreen} the causal (retarded) time-domain Green's function. Further we make some brief remarks on the frequency domain representation the dynamic Green's function representing the solution of the self-similar Helmholtz-equation.

\subsection{Cauchy problem}
\label{cauchy}

The {\it Cauchy problem} is defined as follows: Construct the displacement field $u(x,t)$ solving (\ref{fractlapself})
which we rewrite for our convenience now in the form

\begin{equation}
\label{fractionaleqmov} \frac{\partial^2}{\partial t^2}u(x,t) = -{\cal L}u(x,t)
\end{equation}
with $\Delta_{(\delta,h,\zeta)}=-{\cal L}$ denotes the self-similar Laplacian of (\ref{laplace}).  The operator
${\cal L}$ is necessarily self-adjoint\footnote{i.e. ${\cal L}={\cal L}^+$} and positive definite with ${\cal L}e^{ikx}=\omega^2(k)e^{ikx}$ and the dispersion relation $\omega^2(k) > 0$ for $k\neq 0$ (Eq. (\ref{dispersionrel}) of the last section \ref{deltaforce}). Since ${\cal L}$ is a linear operator we can take advantage of this and employ in the analysis to follow the calculus of linear operators.

The displacement field
$u(x,t)$ is required to fulfill the following initial condition at $t=0$:

\begin{equation}
\label{inidis}
u(x,t=0)= u_0(x)
\end{equation}
with the prescribed initial displacement field $u_0(x)$ at $t=0$. The velocity field $\frac{\partial}{\partial t}u(x,t)$ is required to fulfil the following initial condition at $t=0$:
\begin{equation}
\label{inivel}
\frac{\partial}{\partial t}u(x,t=0)= v_0(x)
\end{equation}
where $v_0(x)$ denotes the prescribed initial velocity field at $t=0$. (\ref{fractionaleqmov}) with the initial conditions (\ref{inidis}) and (\ref{inivel}) represents the Cauchy problem which is defined in the domain $-\infty < x <\infty$ and $-\infty < t < \infty$.
In order to construct
the unique solution of the Cauchy problem, it is convenient to write the displacement field $u(x,t)$ at time $t$ in the form

\begin{equation}
\label{sol1}
u(x,t)=\cos{({\cal L}^{\frac{1}{2}}t)}\, u_0(x)+{\cal L}^{-\frac{1}{2}}\sin{({\cal L}^{\frac{1}{2}}t)}\, v_0(x)
\end{equation}
and as a consequence of (\ref{sol1}) the velocity field $v(x,t)=\frac{\partial}{\partial t}u(x,t)$ at time $t$

\begin{equation}
\label{sol2}
v(x,t)=-{\cal L}^{\frac{1}{2}}\sin{({\cal L}^{\frac{1}{2}}t)}\, u_0(x)+\cos{({\cal L}^{\frac{1}{2}}t)}\, v_0(x)
\end{equation}

Expressions (\ref{sol1}), and (\ref{sol2}) are defined by their power-series involving only entire powers ${\cal L}^n$ with $n=0,1,2,..\in \N_0$ of operator ${\cal L}=-\Delta_{\delta,h,\zeta}$ with the identical operator ${\cal L}^0=1$.

To evaluate expression (\ref{sol1}) it is convenient to write it in its spectral representation which is in our case the Fourier-representation. Introducing the Fourier transformations

\begin{equation}
\label{fou1}
u_0(x)=\frac{1}{2\pi}\int_{-\infty}^{\infty}{\tilde u}_0(k)e^{ikx}{\rm d}k
\end{equation}
and
\begin{equation}
\label{fou2}
v_0(x)=\frac{1}{2\pi}\int_{-\infty}^{\infty}{\tilde v}_0(k)e^{ikx}{\rm d}k
\end{equation}
and taking into account that

\begin{equation}
\label{eigenLhalb}
{\cal L}^{\frac{1}{2}}e^{ikx}=\omega(k)e^{ikx}
\end{equation}
with

\begin{equation}
\label{eigenLhalbfunc}
f({\cal L}^{\frac{1}{2}}t)e^{ikx}=f(\omega(k)t)e^{ikx}
\end{equation}
for any sufficiently smooth function $f(\omega t)=\sum_{m=0}^{\infty}a_m\omega^mt^m$ and where $\omega(k)$ denotes the (positive) square-root of dispersion relation (\ref{dispersionrel}).
It is now convenient to rewrite (\ref{sol1}) in rather trivial manner

\begin{equation}
\label{rewrisol}
u(x,t)=\cos{({\cal L}^{\frac{1}{2}}t)}\, \int_{-\infty}^{\infty}\delta(x-\xi)\,u_0(\xi){\rm d}\xi +{\cal L}^{-\frac{1}{2}}\sin{({\cal L}^{\frac{1}{2}}t)}\,\int_{-\infty}^{\infty}\delta(x-\xi)\,v_0(\xi){\rm d}\xi
\end{equation}
where the operators act only on the dependence on $x$. By using the Fourier representation of the $\delta$-function (\ref{deltafunc}) we have

\begin{equation}
\label{account2}
Q(x,t)={\cal L}^{-\frac{1}{2}}\sin{({\cal L}^{\frac{1}{2}}t)}\delta(x)=
\frac{1}{2\pi}\int_{-\infty}^{\infty}e^{ikx}\,\frac{\sin{(\omega(k)t)}}{\omega(k)}{\rm d}k
\end{equation}
and

\begin{equation}
\label{account1}
\frac{\partial }{\partial t}Q(x,t)=\cos{({\cal L}^{\frac{1}{2}}t)}\delta(x)=
\frac{1}{2\pi}\int_{-\infty}^{\infty}e^{ikx}\,\cos{(\omega(k)t)}{\rm d}k
\end{equation}

We can then write (\ref{rewrisol}) in terms of the kernels $Q(x,t)$ and $\frac{\partial }{\partial t}Q(x,t)$ namely

\begin{equation}
\label{finalrepsol1}
u(x,t)=\int_{-\infty}^{\infty}\frac{\partial }{\partial t}Q(x-\xi,t)u_0(\xi){\rm d}\xi+\int_{-\infty}^{\infty}Q(x-\xi,t)v_0(\xi){\rm d}\xi
\end{equation}
and the velocity field is written as
\begin{equation}
\label{finalrepsol2}
v(x,t)=\int_{-\infty}^{\infty}\frac{\partial^2 }{\partial t^2}Q(x-\xi,t)u_0(\xi){\rm d}\xi+\int_{-\infty}^{\infty}\frac{\partial }{\partial t}Q(x-\xi,t)v_0(\xi){\rm d}\xi
\end{equation}

From relations (\ref{account2}) and (\ref{account1}) we observe the initial conditions

\begin{equation}
\label{iniQ}
Q(x,t=0)=0 \,, \hspace{1cm} \frac{\partial}{\partial t}Q(x,t=0)=\delta(x) \,, \hspace{1cm} \frac{\partial^2}{\partial t^2}Q(x,t=0)=0
\end{equation}
which guarantee that (\ref{finalrepsol1}), (\ref{finalrepsol2}) indeed solve the Cauchy problem.
The problem is solved for our medium by the explicit determination of (\ref{account2}) and (\ref{account1}).
To this end we evaluate relations (\ref{account2}) and (\ref{account1}) which can be written as\footnote{Where the convergence of these series will be verified in appendix \ref{convergence}.}

\begin{equation}
\label{Qxt1}
Q(x,t)= \sum_{n=0}^{\infty}(-1)^n\frac{t^{2n+1}}{(2n+1)!}\, q_n(x)
\end{equation}
and
\begin{equation}
\label{Qxtt2}
\frac{\partial }{\partial t}Q(x,t) = \sum_{n=0}^{\infty}(-1)^n\frac{t^{2n}}{(2n)!}\, q_n(x)
\end{equation}
with the functions $q_n(x)$ defined by

\begin{equation}
\label{qncoeff1}
q_n(x)={\cal L}^n\delta(x)=\frac{1}{2\pi}\int_{-\infty}^{\infty}e^{ikx}\,\omega^{2n}(k)\,{\rm d}k=:\lim_{\epsilon \rightarrow 0+}\frac{1}{\pi}
\int_{0}^{\infty}e^{-k\epsilon}\omega^{2n}(k) \cos{kx}\,{\rm d}k \,, n=0,1,2,..\in \N_0
\end{equation}
The integral $\int_{0}^{\infty}e^{-k\epsilon}k^{\alpha}\cos{kx}\,{\rm d}k $
exists for $\alpha >-1$ and contains the entire admissible range $0<\delta<2$ of the Laplacian (\ref{laplace}).
Hence we have

\begin{equation}
\label{qncoeff2}
q_n(x)=\lim_{\epsilon \rightarrow 0+}\frac{A_{\delta}^n}{\pi}\int_{0}^{\infty}e^{-k\epsilon}k^{\delta n} \cos{kx}\,{\rm d}k=\frac{A_{\delta}^n}{\pi}Re(J_{n\delta}) \,, n=0,1,2,..\in \N_0\,,0<\delta<2
\end{equation}
with

\begin{equation}
\label{Jndelta}
J_{n\delta}=\lim_{\epsilon\rightarrow 0+}\frac{1}{{(\epsilon - i|x|)}^{(n\delta+1)}}\int_0^{\infty}e^{-s}s^{n\delta}{\rm d }s=\lim_{\epsilon\rightarrow 0+}\frac{(n\delta)!}{{(\epsilon - i|x|)}^{(n\delta+1)}}
\left\{\begin{array}{l}
\ds i\, e^{i\frac{\pi \delta n}{2}}\, \frac{(n\delta)!}{|x|^{n\delta + 1}} \,,x\neq 0\,,\hspace{0.2cm} n=1,2,..\in \N \nonumber \\ \nonumber \\
\nonumber \\ \ds \lim_{\epsilon\rightarrow 0+}\frac{1}{\epsilon - i|x|}=\pi\delta(x)+i{\cal P}(\frac{1}{x})\,,\hspace{0.2cm} n=0
\end{array}
\right.
\end{equation}
where $(n\delta)!=\Gamma(n\delta+1)$ and ${\cal P}(..)$ denotes the principal value of $(..)$. We obtain then
\begin{equation}
\label{integtype2}
q_n(x)=q_n(|x|)=\left\{\begin{array}{l}\frac{A_{\delta}^n}{\pi} Re(J_{n\delta})=-\frac{A_{\delta}^n}{\pi}\sin{(\frac{\pi \delta n}{2})}\,\frac{(n\delta)!}{|x|^{n\delta + 1}} \,,\hspace{0.5cm}x\neq 0\,, n=1,2,..\in\N\,,\hspace{0.2cm} 0 < \delta < 2 \nonumber \\ \nonumber \\
q_0(x)=\frac{1}{\pi} Re(J_{0})=\delta(x) = 0 \,,\hspace{0.5cm}x\neq 0
\end{array}
\right.
\end{equation}
where $A_{\delta}$ is determined in (\ref{Adelta2}). On the right-hand sides of (\ref{Jndelta}), (\ref{integtype2}) we skipped a singular term which also occurs in the case $n>0$ at $x=0$. Therefore this relation excludes $x=0$ for $n>0$. We come back to this important point in section \ref{selfsimpot} by considering integrals of the $J_{\alpha}$ for the entire range of their existence ($\alpha >-1$).

The series (\ref{Qxt1}), (\ref{Qxtt2}) together with functions (\ref{integtype2}) constitute the solution of the Cauchy problem being defined in the entire range $0<\delta<2$ of existence of Laplacian (\ref{laplace}).
It is shown in appendix \ref{convergence} that the series for (\ref{Qxt1}), (\ref{Qxtt2}) converge uniformly absolutely for all $x$ and $t$.

\subsection{Retarded time domain Green's function}
\label{dyngreen}

In view of the results of the last section it is only a small step to determine
the {\it retarded} (or causal) dynamic Green's function. The dynamic Green's function is defined as the solution of the self-similar wave-equation

\begin{equation}
\label{dygrefu}
\left((\frac{\partial}{\partial t}+\epsilon)^2-\Delta_{(\delta,h,\zeta)}\right)g(x,t)=\delta(x)\delta(t)
\end{equation}
with the self-similar Laplacian $\Delta_{(\delta,h,\zeta)}$ defined by (\ref{laplace}).
{\it Causality} means that $g(x,t)$ must be zero before and nonzero only after the impact of the pulse $\delta(x)\delta(t)$ i.e. $g(x,t<0)=0$ and $g(x,t>0)\neq 0$ \cite{jackson,michel2}.
We introduced in (\ref{dygrefu}) an infinitesimal {\it positive} damping term $\epsilon\rightarrow 0+$ which breaks the time-inversion symmetry of the wave operator and guarantees in this way
$g(x,t)$ being the {\it causal} solution of (\ref{dygrefu}).
The physical interpretation of $g$ is the displacement field due to a $\delta$-type force density
of the form of the right hand side of (\ref{dygrefu}). The Green's function defined by
(\ref{dygrefu}) solves the dynamic problem

\begin{equation}
\label{dynaprob}
\left((\frac{\partial}{\partial t}+\epsilon)^2-\Delta_{(\delta,h,\zeta)}\right)u(x,t)=f(x,t)
\end{equation}
where $f(x,t)$ is the density of external forces and $u(x,t)$ the corresponding displacement field
by the convolution

\begin{equation}
\label{greenconv}
u(x,t)=\int_{-\infty}^{\infty}\int_{-\infty}^{t}g(x-\xi,t-\tau)f(\xi,\tau)\,{\rm d}\xi\,{\rm d}\tau
\end{equation}
where $g(x,t)$ indicates the Green's function defined by (\ref{dygrefu}). The upper limit
in the time-integration in (\ref{greenconv}) is due to the causality of $g(x,t)$.
With the Fourier-transformation

\begin{equation}
\label{spacefourier1}
g(x,t)=\frac{1}{2\pi}\int_{-\infty}^{\infty}{\tilde g}(k,t)e^{ikx}\,{\rm d}k
\end{equation}
and
\begin{equation}
\label{spacefourier2deltafu}
\delta(x)=\frac{1}{2\pi}\int_{-\infty}^{\infty}e^{ikx}\,{\rm d}k
\end{equation}
yields (\ref{dygrefu}) the evolution equation for the normal amplitudes $g(k,t)$
in the form

\begin{equation}
\label{harmosc}
\left((\frac{\partial}{\partial t}+\epsilon)^2+\omega^2(k)\right){\tilde g}(k,t)=\delta(t)
\end{equation}
which is nothing but the equation for the causal Green's function of a damped harmonic oscillator
with damping constant\footnote{$\epsilon\rightarrow 0+$ being positive and infinitesimal being sufficient to obtain the causal solution.} $\epsilon>0$ and eigenfrequency $\omega(k)$ which has the (causal) solution of the well-known form\footnote{See any textbook of theoretical physics or e.g. \cite{michel2}.}

\begin{equation}
\label{causalsol1}
{\tilde g}(k,t)=e^{-\epsilon t}\Theta(t)\frac{\sin{\omega(k)t}}{\omega(k)}
\end{equation}
where $\Theta(t)$ denotes the Heaviside unit-step function being defined as $\Theta(t)=1$ for $t>0$ and $\Theta(t)=0$ for $t<0$. (\ref{causalsol1}) holds for any (also non-infinitesimal) damping $\epsilon >0$ where the dissipative factor $e^{-\epsilon t}$ can be skipped in the limit of an infinitesimal $\epsilon>0$. The frequency $\omega(k)=\sqrt{A_{\delta}}|k|^{\frac{\delta}{2}}$ denotes the positive root of
the dispersion relation (\ref{dispersionrel}) holding for $0<\delta<2$.
In view of the relation
\begin{equation}
\label{auxiliary}
{\cal L}^{-\frac{1}{2}}\sin{{\cal L}^{\frac{1}{2}}t}\,e^{ikx}=\frac{\sin{\omega(k)t}}{\omega(k)}e^{ikx}
\end{equation}
with ${\cal L}=-\Delta_{\delta,h,\zeta}$ we have the retarded space-time Green's function $g(x,t)$ in the form

\begin{equation}
\label{ggreenret}
g(x,t)=e^{-\epsilon t}\,\Theta(t){\cal L}^{-\frac{1}{2}}\,\sin{({\cal L}^{\frac{1}{2}}t)}\,\delta(x)=e^{-\epsilon t}\,\Theta(t)\,Q(x,t)
\end{equation}
where $Q(x,t)$ has been evaluated in relations (\ref{Qxt1}) ff.
\newline\newline
\noindent {\bf Some remarks on the dynamic Green's function in the space-frequency domain}
\newline
For the sake of completeness we give a brief representation of the frequency-space Green's function, i.e.  Green's function of the self-similar Helmholtz equation. The Helmholtz Green's function is defined as
\begin{equation}
\label{helmholtzgf}
{\hat g}(x,\omega)=\int_{0}^{\infty}e^{i\omega t}g(x,t)\,{\rm d}t
\end{equation}
where $g(x,t)$ is the retarded space-time-domain Green's function (\ref{ggreenret}) and the lower integration limit is due to causality. (\ref{helmholtzgf}) solves the self-similar Helmholtz-equation

\begin{equation}
\label{selfsimhelmholtz}
\left((\omega+i\epsilon)^2+\Delta_{(\delta,h,\zeta)}\right){\hat g}(x,\omega)=-\delta(x)
\end{equation}
where $\epsilon >0$ is a positive infinitesimal damping constant which guarantees existence of the Helmholtz Green's operator defined in (\ref{helmopinvert}) below and causality in the time domain.
We can invert this Helmholtz equation with $\Delta_{(\delta,h,\zeta)}=-{\cal L}$ where the Helmholtz Green's function can be written as

\begin{equation}
\label{helmopinvert}
{\hat g}(x,\omega)=\left({\cal L}-(\omega+i\epsilon)^2\right)^{-1}\delta(x)
\end{equation}
with the Green's operator (resolvent operator) of the frequency-domain $\left({\cal L}-(\omega+i\epsilon)^2\right)^{-1}$.

\section{Self-similar diffusion problem - L\'evi flights}
\label{transport}
This section is devoted to the analysis of the diffusion problem governed by the self-similar Laplacian (\ref{laplace}). We consider an ensemble of particles with local concentration $\rho(x,t)$. We also can conceive this problem to be defined below as a fractional differential equation for a (probability)-distribution $\rho(x,t)$. The idea to represent statistical distributions as solutions of fractional differential equations is actually not new, we refer for instance to the paper of Li et al. \cite{Li} where "stable" distributions are generated by fractional differential equations. However, so far the fractional differential equations generating statistical distributions as solutions were rather a kind of "guesswork" without having an intuitive physical "picture" about the generating fractional differential equation. The goal of this section is to analyze a diffusion-equation employing the self-similar Laplacian (\ref{laplace}) of the form

\begin{equation}
\label{diffussioneq1}
\frac{\partial }{\partial t}\rho(x,t)=-{\cal L}\rho(x,t)
\end{equation}
where $-{\cal L}=\Delta_{(\delta,h,\epsilon)}$ indicates the self-similar Laplacian (\ref{laplace}).
We consider this problem in the domain $-\infty < x < \infty$ and $t>0$. It is important that we restrict on $t>0$ as diffussion is a time-irreversible process\footnote{That means the time-inverse process is physically inadmissible.} which is expressed by the odd-order of the time derivative in (\ref{diffussioneq1}).
$\rho(x,t)$ denotes the density of the diffusing particles but can be also conceived as
a probability-density distribution-function.
(\ref{diffussioneq1}) recovers the traditional diffusion equation by replacing the self-similar Laplacian $\Delta_{\delta,h,\zeta}$ by the traditional 1D-Laplacian $\frac{\partial^2}{\partial x^2}$ leading to a gaussian distribution.

We give an intuitive physical picture of the diffusion processes described by (\ref{diffussioneq1}) as follows:
Let us assume we have $\rho(x,t){\rm d}x$ particles at time $t$ at location $x$. Any particle is allowed to jump at time $t$ from position $x$ to arrive at $t+dt$ at any location $x\pm \tau$ different from $x$ (with $0<\tau<\infty$). The jump-rate of this process is measured by
$(\rho(x+\tau,t)+\rho(x-\tau,t)-2\rho(x,t))/\tau^{-(1+\delta)}$ which is given by the integrand of (\ref{laplace}) where $0<\delta< 2$. We observe the following: The jump-rate from $x$ to $x\pm \tau$ is zero if
$(\rho(x+\tau,t)+\rho(x-\tau,t)-2\rho(x,t))=0$, i.e. there are no jumps for a equal distribution $\rho=const$. Jumps to "far" locations are more seldom as jumps to "close" locations for equal $(\rho(x+\tau,t)+\rho(x-\tau,t)-2\rho(x,t))\neq 0$.
The possibility of a particle to make jumps of arbitrary distances is expressed by the non-locality and continuity of Laplacian (\ref{laplace}).
Particle-jumps to any location are possible in (\ref{diffussioneq1}), whereas non-local jumps are suppressed in the traditional (gaussian) case which is reflected by the locality
of the traditional Laplacian $\frac{\partial^2}{\partial x^2}$.
The non-local Laplacian (\ref{laplace}) sums up over all possible jump-events by which particles can reach to or escape from a space-point $x$.
For $\rho=const$ the jumping rate is zero (since ${\cal L}1=0$) and, as we will show, the equal-distribution is a stationary solution and $\rho=0$ an {\it attractor} for $t\rightarrow \infty$ for any initial distribution $\rho_0(x)$ just as in the case of gaussian distribution.

If we assume $\rho(x,t)$ to be the probability distribution of a single particle and $\rho(x,t){\rm d}x$ to be the probability to find a particle in the interval within $[x,x+{\rm d}x]$ at time $t$,
then the trajectory of the particle propagating due to probability-distribution $\rho(x,t)$ which evolves according to
(\ref{diffussioneq1}) is discontinuous and erratic whereas in the gaussian case the particle-motion describes a continuous trajectory. The discontinuous characteristics of the particle trajectory associated with (\ref{diffussioneq1}) becomes more and more pronounced as $\delta\rightarrow 0$ which is due to the weaker decay of the kernel $\tau^{-{(\delta+1)}}$ for $\tau\rightarrow\infty$. We should like to mention that motions characterized by jump-probability distributions of the form of a power law as $\tau^{-{(\delta+1)}}$ are known in the literature as {\it L\'evi-flights} where $\delta$ is called L\'evi-parameter. L\'evi flights correspond to L\'evi-stable probability distributions for $0<\delta\leq 2$ \cite{mandel2}.
Typical features of L\'evi-distributions are their infinite variances in the range $0<\delta< 2$. The present analysis will end up in probability-distributions of exactly this type.

In order to define diffusion problem (\ref{diffussioneq1}) as initial value problem, we prescribe an initial distribution $\rho_0(x)$ at $t=0$ as

\begin{equation}
\label{inidis1}
\rho(x,t=0)=\rho_0(x)
\end{equation}

We assume that the particle number is conserved in time, there is neither particle generation nor annihilation. The diffusion equation (\ref{diffussioneq1}) describes then the continuity equation accounting for the particle (probability)
balance at space-point $x$ and time $t$ having the form

\begin{equation}
 \label{continuity}
\frac{\partial }{\partial t}\rho(x,t)+ \nabla\cdot j(x,t)=0
\end{equation}
where $\nabla=\frac{\partial}{\partial x}$ denotes the usual 1D gradient operator.
Unlike in the Gaussian case where the traditional Laplacian comes into play by assuming Fick's law as a consequence of a gradient dynamics as constitutive law, in our case the constitutive law is instead of a (local) gradient a non-local constitutive law for the particle (probability) flux. An explicit form of the
particle (probability) flux is obtained in view of (\ref{laplaceb}) as

\begin{equation}
\label{laplacebflux}
j(x,t)=-\frac{h^{\delta}}{\zeta\delta}\int_0^{\infty}\frac{\left(\rho(x+\tau,t)-\rho(x-\tau,t)\right)}{\tau^{\delta}}\,{\rm
d}\tau \,,\hspace{2cm} 0<\delta<2
\end{equation}
which is our self-similar constitutive law and can be conceived as the self-similar analogue to Fick's law $j_{gauss}(x,t)=-\nabla \rho(x,t)$ of the Gaussian case. In (\ref{laplacebflux}) an irrelevant integration-constant has been neglected. We can consider (\ref{laplacebflux})
as definition of the self-adjoint part of the self-similar gradient operator acting on $\rho(x,t)$.
The distribution function $\rho(x,t)$ is required to fulfil the normalization condition

\begin{equation}
\label{normalization1}
\int_{-\infty}^{\infty}\rho(x,t){\rm d}x=\int_{-\infty}^{\infty}\rho_0(x){\rm d}x=1 \,,\hspace{1.5cm} \forall t >0
\end{equation}

(\ref{normalization1}) indicates that $\rho(x,t)$ has to be a normalized {\it distribution}.
$\rho(x,t)$ is uniquely determined by
(\ref{diffussioneq1}) and (\ref{inidis1}) and can be written in the form

\begin{equation}
\label{expsol}
\rho(x,t)= e^{-{\cal L}t}\rho_0(x)
\end{equation}
where we restrict us to $t>0$.
The question to be answered in the following is: Does (\ref{expsol}) represent a normalized distribution, i.e. fulfills (\ref{normalization1}) under the condition that $\rho_0(x)$ represents a normalized distribution? We can answer this question by "yes" by giving the following formal prove by integrating (\ref{expsol})

\begin{equation}
\label{expsol2}
\int_{-\infty}^{\infty}\rho(x,t)\,{\rm d}x= e^{-{\cal L}t}\int_{-\infty}^{\infty}\rho_0(x)\,{\rm d}x= e^{-{\cal L}t}\,1=1
\end{equation}
where we take into account that ${\cal L}^n1=\delta_{n0}$ since Laplacian (\ref{laplace}) applied to a constant yields zero.
Let evaluate (\ref{expsol}) and deduce the propagator (kernel) that represents $\rho(x,t)$ in terms of a convolution

\begin{equation}
\label{convoldiss1}
\rho(x,t)=\int_{-\infty}^{\infty}W(x-\xi,t)\rho_0(\xi){\rm d}\xi=e^{-{\cal L}t}\int_{-\infty}^{\infty}\delta(x-\xi)\rho_0(\xi)\,{\rm d}\xi
\end{equation}

The propagator $W(x,t)$ is obtained by
\begin{equation}
\label{propag}
W(x,t)= e^{-{\cal L}t}\delta(x)
\end{equation}
and fulfils the initial condition
\begin{equation}
\label{iniWcon}
W(x,t=0)=\delta(x)
\end{equation}
The necessary and sufficient condition that $W(x,t)$ describes a normalized probability distribution is

\begin{equation}
\label{normcond}
\int_{-\infty}^{\infty} W(x,t)\,{\rm d}x=1
\end{equation}
which is indeed fulfilled as can be seen by putting $\rho_0(x)=\delta(x)$ in (\ref{expsol2}).
If $\rho(x,t)$ is conceived as a probability-distribution $W(x,t)$ is the
{\it conditional probability density} to find a particle which was at $t=0$ located at
$x=0$ at time $t>0$ at point $x$.
The propagator $W(x,t)$ can be evaluated by using ${\cal L}e^{ikx}=\omega^2(k)e^{ikx}$
in the form
\begin{equation}
\label{propa1}
W(x,t)= e^{-{\cal L}t}\delta(x)=\frac{1}{2\pi}\int_{-\infty}^{\infty}e^{ikx}e^{-\omega^2(k)t}\,{\rm d}k=\frac{1}{\pi}\int_0^{\infty}e^{-A_{\delta}k^{\delta}t}\cos{(kx)}\, {\rm d}k
\end{equation}
where $\omega^2(k)=A|k|^{\delta}$ is the dispersion relation
(\ref{dispersionrel}) which is defined in $0<\delta< 2$. Integral (\ref{propa1}) exists in the whole range
of $0<\delta<2$ where $A_{\delta}>0$ for $t>0$. 
We observe further because of $\omega^2(k)\geq 0$ that
$e^{-\omega^2(k)t} \rightarrow 0$ at $t\rightarrow\infty$ thus (\ref{propa1}) tends to zero
as $t$ tends to infinity

\begin{equation}
\label{Watract}
\lim_{t\rightarrow\infty} W(x,t) = 0
\end{equation}
so the stationary solution $\rho_{\infty}=0$ is always an attractor independent of the initial distribution $\rho_0(x)$ in the physical admissible interval $0<\delta<2$. It is important to note that for any {\it finite} $0<t<\infty$ (\ref{Watract}) maintains in some regions non-zero (positive) values
in order to fulfil (\ref{normcond}). The goal is now to evaluate (\ref{propa1}).

It can be further seen in (\ref{propa1}) since $e^{-A_{\delta}k^{\delta}t}$ is a monotonous
decreasing function in $k$ that the positive contributions due to $\cos{kx}$ dominate
the negative contributions in the last integral of (\ref{propa1}) so that $W(x,t)$ fulfills another necessary characteristics of a probability-distribution, namely
\begin{equation}
\label{posW}
W(x,t) \geq 0 \hspace{1cm} \forall x \in \R \hspace{0.3cm} {\rm and}\hspace{0.2cm} t>0
\end{equation}
So we can infer that the initial value problem defined by (\ref{diffussioneq1}), (\ref{inidis1}) with the self-similar Laplacian (\ref{laplace}) indeed
describes a diffusion problem (transport problem) under the condition that (\ref{inidis1}) is a normalized distribution.
It remains us now to evaluate (\ref{propa1}) in more explicit form
\begin{equation}
\label{propa2}
W(x,t)=\frac{1}{\pi} Re\left\{\int_0^{\infty}e^{ikx-A_{\delta}k^{\delta}t}\, {\rm d}k\right\}\,,\hspace{2cm} 0<\delta<2
\end{equation}
where $0<\delta<2$ is the range of existence of $A_{\delta}$.
Distributions of the form (\ref{propa1}), (\ref{propa2}) have already been proposed by L\'evi \cite{mandel1,levi} and refers to the category of a "L\'evi-stable" or "L-stable" distribution
(for further details on L-stable distribution we refer e.g. to \cite{mandel1}, p. 106ff. and the references therein).
For $\delta=1$ we can directly evaluate (\ref{propa2}) directly and obtain

\begin{equation}
\label{propa2delta1}
W_{\delta=1}(x,t)= Re\left\{ \frac{1}{\pi}\int_0^{\infty}e^{-k(A_1t-ix)}\, {\rm d}k\right\}
\end{equation}
which yields

\begin{equation}
\label{propa2cauchy}
W_{\delta=1}(x,t)=\frac{1}{\pi}\frac{A_1t}{(x^2+(A_1t)^2)}
\end{equation}
where $A_1=\frac{h\pi}{\zeta}$. Relation (\ref{propa2cauchy}) represents spatially a distribution of the Cauchy type with $A_1t$ being a "parameter". One verifies directly that (\ref{propa2cauchy}) fulfills the conditions (\ref{iniWcon}), (\ref{normcond}) and (\ref{Watract}).
L\'evi already gave static spatial probability-distributions of the general form (\ref{propa2}) \cite{mandel2} and found that in general they are admissible for $0<\delta\leq 2$ where $\delta=1$ corresponds to the Cauchy-distribution and $\delta=2$
to the Gaussian distribution.
For $\delta=2$ the Laplcian (\ref{laplace}) does not exist and hence $A_2$ is not defined in our approach.

Mandelbrot showed that L-stable distributions play a crucial role in the description of economic processes and stock courses and describe the statistics of fractal irregular trajectories.
The irregularity of these trajectories decreases from $\delta=0$ to $\delta\rightarrow 2$ \cite{mandel2} which is consistent with our above given physical picture. The irregularity is due to the fact that (\ref{propa2}) produces in the admissible range of existence of the Laplacian (\ref{laplace}) $0<\delta<2$ infinite fluctuations due to discontinuous trajectories. The Cauchy distribution ($\delta=1$) itself refers also to this category. Only the Gaussian case corresponds to a diffusion equation with a traditional Laplacian describing continuous trajectories (Brownian motion) with {\it finite} mean fluctuation (variance). A consequence is that the unjustified use of Gaussian statistics underestimates fluctuations and risks! We refer in this context to \cite{mandel,mandel1,mandel2,sapoval} and the references therein.
We will reconsider briefly this important point at the end of this section.

It still remains to evaluate (\ref{propa2}) which we do by expanding $e^{-\omega^2(k)t}$ in a series and arrive at the expression

\begin{equation}
\label{propa3}
W(x,t)= \sum_{n=0}^{\infty}(-1)^n\frac{t^n}{n!}\,q_n(x) \,,\hspace{2cm} 0<\delta<1
\end{equation}
where this series converges only in the range $0<\delta<1$ which is shown in appendix \ref{conv2}. In (\ref{propa3}) appear the functions
\begin{equation}
\label{qnencore}
q_n(x)={\cal L}^n\delta(x)=\frac{1}{2\pi}\int_{-\infty}^{\infty}e^{ikx}\,\omega^{2n}(k)\,{\rm d}k
\end{equation}
which where already determined in the last section by (\ref{qncoeff1})-(\ref{integtype2}). Due to their importance for our approach we devote the next section \ref{selfsimpot} to their thorough analysis.
The series (\ref{propa3}) writes

\begin{equation}
\label{Wsingular}
W(x,t)=\frac{1}{\pi}\lim_{\epsilon\rightarrow 0+}Re\left\{\sum_{n=0}^{\infty}
\frac{(-1)^n((\delta n)!(A_{\delta}t)^n)}{n!(\epsilon -i|x|)^{n\delta+1}}\right\}=\delta(x)+\frac{1}{\pi}\lim_{\epsilon\rightarrow 0+}Re\left\{\sum_{n=1}^{\infty}
\frac{(-1)^n((\delta n)!)(A_{\delta}t)^n}{n!(\epsilon -i|x|)^{n\delta+1}}\right\}
\end{equation}
which assumes for $x\neq 0$
\begin{equation}
\label{wfunction2}
W(x,t)=\frac{1}{\pi}\sum_{n=1}^{\infty}(-1)^{n-1}\frac{(n\delta)!}{n!}\sin{(\frac{\pi n\delta}{2})} {\frac{A_{\delta}^nt^n}{|x|^{n\delta+1}}} \,,\hspace{2cm} 0<\delta<1
\end{equation}
where the singular parts at $x=0$ vanishing for $x\neq 0$ have been omitted. (\ref{wfunction2}) has the form
\begin{equation}
\label{explicitW}
W(x,t)=-\frac{1}{\pi|x|}Im\{w(\xi(x,t))\}
\end{equation}
when we introduce the variable

\begin{equation}
\label{coordsup}
\xi(x,t)=\frac{t}{|x|^{\delta}}e^{\frac{{\pi i \delta}}{2}} A_{\delta}
\end{equation}
with the function $w(\xi)$

\begin{equation}
\label{wfunction}
w(\xi)=\sum_{n=1}^{\infty}(-1)^n\frac{(n\delta)!}{n!}\xi^n \,,\hspace{2cm} 0<\delta<1
\end{equation}
where series representations (\ref{propa3}), (\ref{Wsingular}) converge $\forall \xi$ in the interval $0<\delta<1$ (appendix \ref{conv2}).

Let us now discuss some of the main characteristics of the distribution $W(x,t)$ which is defined in the whole range $0<\delta<2$: In view of (\ref{propa1}) and (\ref{integtype2}) we observe that $W(x,t)=W(-x,t)=W(|x|,t)$ is a symmetric distribution
with respect to $x$ and as a consequence all moments of odd order

\begin{equation}
\label{oddmo}
<x^{2n+1}>=\int_{-\infty}^{\infty}W(x,t)\, x^{2n+1}\,{\rm d}x=0 \,,\hspace{2cm} n=0,1,..\in \N_0
\end{equation}
are vanishing especially the mean-value $<x>=0$. The moments of even order are obtained
from
\begin{equation}
\label{evenmo}
<x^{2n}>=\int_{-\infty}^{\infty}W(x,t)\, x^{2n}\,{\rm d}x\rightarrow \infty \,,\hspace{2cm} n=1,..\in \N
\end{equation}
are diverging.
We can conclude without any further calculation that (\ref{evenmo}) is divergent since any {\it real valued power
function} (except a constant) is not admissible in (\ref{laplace}). The existence of (\ref{laplace}) however is achieved for certain
complex valued power functions as we show in the next section \ref{selfsimpot}.

\section{Self-similar potentials}
\label{selfsimpot}

Finally we  analyze more closely functions of the type $q_n(x)$ of (\ref{qnencore}) which were determined in (\ref{qncoeff1})-(\ref{integtype2}). To this end let us evoke the definition of the $q_n(x)={\cal L}^n\delta(x)$ which was defined by using the Laplacian (\ref{laplace}), (\ref{laplaceb}). We should emphasize that $\Delta_{(\delta,h,\zeta)}=-{\cal L}$ depends on the continuous parameter $\delta$ which was
restricted in (\ref{laplace}), (\ref{laplaceb}) to the interval $0<\delta<2$ where the upper limit $\delta<2$
came into play since $2u(x)-u(x+\tau)-u(x-\tau)\rightarrow \, C(x) \tau^2$ is a quadratic function of $\tau$ for small $\tau$. In this subsection we will show that for a certain class of {\it singular} functions, namely functions of the form $b_{n\delta}(x)={\cal L}^n\delta(x)/A_{\delta}^n$, the upper limit $\delta < 2$ {\it does not exist} and for which the condition $\delta >0$ is sufficient.
We call these functions due to their importance as solutions of the Laplace-equation (\ref{laplace}) as "self-similar potentials".
Their singular behavior at $x=0$ becomes especially important when we consider integrals or convolutions of these functions. Whereas any {\it real-valued power-functions} $x^{\beta}$ are not admissible in (\ref{laplace}). Self-similar potentials to be analyzed, however, are admissible and give the only way  to an admissible definition of powers ${\cal L}^n$.
Let us consider the Fourier transformation of $|k|^{\alpha}e^{-|k|\epsilon}$ with $\epsilon >0$, namely\footnote{For further details on the Fourier transform of $|k|^{\alpha}$, see \cite{gelfand}}

\begin{equation}
\label{intalph}
b_{\alpha}(x)= \lim_{\epsilon \rightarrow 0+}\frac{1}{2\pi}\int_{-\infty}^{\infty}e^{ikx-|k|\epsilon}|k|^{\alpha}{\rm d}k=\lim_{\epsilon \rightarrow 0+}\frac{1}{\pi}\int_0^{\infty}e^{-\epsilon k}k^{\alpha}\cos{kx}\,{\rm d}k \,,\hspace{2cm} \alpha > -1 \in \R
\end{equation}
which exists for $\alpha >-1$.
This function can be identified with the $q_n(x)$ of (\ref{qnencore}) for $\alpha=n\delta$.
We can rewrite (\ref{intalph}) by introducing the integration variable $s=k(\epsilon - i|x|)$ in the form
\begin{equation}
\label{intalpha}
b_{\alpha}(x)= Re\left\{\lim_{\epsilon \rightarrow 0+}\frac{1}{\pi}\frac{1}{(\epsilon-i|x|)^{\alpha+1}}\int_0^{\infty} e^{-s}s^{\alpha}{\rm d }s
\right\}=\frac{\alpha !}{\pi}Re\left\{\lim_{\epsilon \rightarrow 0+}\frac{i^{\alpha+1}}{(x+i\epsilon)^{\alpha+1}} \right\} \,,\hspace{2cm} \alpha > -1 \in \R
\end{equation}
which holds for $\alpha > -1$ and where $\alpha ! =\Gamma(\alpha+1)$. We note that $(x+i\epsilon)^{-\alpha-1}$ assumes the complex conjugate value when replacing $x \leftrightarrow -x$. So its real-part depends only on $|x|$ and we can rewrite

\begin{equation}
 \label{balphre}
b_{\alpha}(x)=b_{\alpha}(|x|)=\frac{\alpha !}{\pi}Re\left\{\lim_{\epsilon \rightarrow 0+}\frac{i^{\alpha+1}}{(|x|+i\epsilon)^{\alpha+1}} \right\} \,,\hspace{2cm} \alpha > -1 \in \R
\end{equation}

Expressions (\ref{intalpha}), (\ref{balphre}) are in accordance with with the result given by
Gel'fand and Shilov for the Fourier transform of $|k|^{\alpha}$ (page 447, equation (13) in \cite{gelfand}).
\newline\newline
\noindent We summarize the following important observations:
\newline\newline
\noindent{\bf (i)} $b_{\alpha}(x)=b_{\alpha}(-x)=b_{\alpha}(|x|)$ is a symmetric function in $x$ for all $\alpha > -1$.
\newline\newline
\noindent{\bf (ii)} $b_{\alpha=0}(x)=\delta(x)$ represents the usual Dirac-$\delta$-function, i.e. is {\it localized} at $x=0$.
\newline\newline
\noindent{\bf (iii)} For $\alpha=2n$ with $n=0,1,2,..\in \N_0$ (\ref{intalpha}) takes the form
of even derivatives of the $\delta$-function
\begin{equation}
\label{int2na}
b_{2n}(x)=(-1)^n\frac{d^{2n}}{d x^{2n}}\delta(x)
\end{equation}
which are {\it localized} at $x=0$.
\newline\newline
\noindent{\bf (iv)} $\alpha>-1 \in \R$:
\newline For $x\neq 0$ (\ref{intalpha}), $b_{\alpha}$ has the explicit form (for $x\neq 0$ we can put directly $\epsilon=0$ in (\ref{intalpha}))

\begin{equation}
\label{bexplaalph}
b_{\alpha}(x)=-\frac{\alpha !}{\pi|x|^{\alpha+1}}\sin{(\frac{\alpha\pi}{2})}
\end{equation}
where we directly verify that this expression is zero for $\alpha=2n$ (in accordance with case (iii) for $x\neq 0$).
For $\alpha>-1 \in \R \notin 0,2,4,..$ expression (\ref{bexplaalph}) is non-zero for
$x\neq 0$ and hence in contrast to the Dirac's $\delta$-function, {\it non-local}. Expression (\ref{bexplaalph}) includes also the
case of odd integers $\alpha=\alpha_n=2n+1$ where $n=0,1,2,..\in \N_0$.

We observe that $b_{\alpha}(x)$ defined by (\ref{intalph}) can be conceived as the Fourier transform of $|k|^{\alpha}$ which exists only for $\alpha>-1$. For exponents $\alpha < -1$ the integral (\ref{intalph}) diverges. However, we can define a {\it regularized Fourier
transform} in the spirit of generalized functions as given in Gel'fand and Shilov \cite{gelfand}) which exists also for $\alpha < -1$. We devote appendix \ref{balphamin1} to this case.
\newline\newline
Let us consider now the following integral
\begin{equation}
\label{bexplaalphint}
\int_{-\infty}^{\infty}b_{\alpha}(x){\rm d}x =\int_{-\infty}^{\infty} e^{-|k|\epsilon}|k|^{\alpha}\delta(k){\rm d}k =
\left\{\begin{array}{ll}
    0 \,, &\alpha > 0 \nonumber \\
 1 \,,  &\alpha=0   \nonumber \\
\infty \,, &-1<\alpha<0
\end{array}\right.
\end{equation}
\newline\newline
\noindent{\bf (v)}
For $\alpha>0$ we observe that integral (\ref{bexplaalphint}) is vanishing.
On the other hand we observe that $b_{\alpha}(x)\neq 0$ for $x\neq 0$ and $sign(b_{\alpha}(x))=-sign(\sin{(\frac{\alpha\pi}{2})})$ is the {\it the same} $\forall x \neq 0$.
\newline
Let us raise the following question: How is it possible that the integral (\ref{bexplaalphint}) is vanishing for $\alpha >0$ whereas $b_{\alpha}\neq 0$
having the {\it same sign} as $-sign(\sin{(\frac{\alpha\pi}{2})})$ almost everywhere, i.e. $\forall x\neq 0$?
To give the answer, let us consider the integral

\begin{equation}
\label{intainf}
{\cal I}_{\alpha}(a)=\int_{a>0}^{\infty}b_{\alpha}(x){\rm d}x = -\frac{(\alpha-1)!}{\pi a^{\alpha}}\sin{(\frac{\pi \alpha}{2})} \,,\hspace{2cm} \alpha >0
\end{equation}
where we have the property $\lim_{a\rightarrow 0+}|{\cal I}_{\alpha}(a)|\rightarrow \infty$.
Let us again ask, why does (\ref{intainf}) not converge in the limiting case $a\rightarrow 0+$ 
towards zero for $\alpha >0$ as indicated in (\ref{bexplaalphint}) since $b_ {\alpha}(x)=b_{\alpha}(-x)$ is an even function? If integral (\ref{bexplaalphint}) indeed is vanishing, then (\ref{intainf}) should be compensated by the integral

\begin{equation}
\label{intainfcomp}
{\cal J}_{\alpha}(a)=\int_{0}^{a}b_{\alpha}(x){\rm d}x =\frac{1}{2}\int_{-a}^{a}b_{\alpha}(x){\rm d}x  \,,\hspace{2cm} \alpha >0
\end{equation}
since $\int_{0}^{a}b_{\alpha}(x){\rm d}x + \int_{a}^{\infty}b_{\alpha}(x){\rm d}x =
\frac{1}{2}\int_{-\infty}^{\infty}b_{\alpha}(x){\rm d}x =0$ ($b(x)=b(|x|)$).

To evaluate this integral we have to use (\ref{intalph}) or directly (\ref{balphre}) which hold $\forall x$ {\it including} $x=0$ to arrive at

\begin{equation}
\label{evintzeroa}
{\cal J}_{\alpha}(a)=\lim_{\epsilon \rightarrow 0+}\frac{1}{\pi}\int_0^{\infty}e^{-\epsilon k}k^{\alpha}\left(\int_0^a\cos{kx}{\rm d}x\right){\rm d}k
\end{equation}
which yields
\begin{equation}
\label{evintyields}
{\cal J}_{\alpha}(a)=\lim_{\epsilon \rightarrow 0+}\frac{1}{\pi}\int_0^{\infty}e^{-\epsilon k}k^{\alpha-1}\sin{(|k|a)}\,{\rm d}k=\frac{(\alpha-1)!}{\pi}Im\left\{\frac{1}{(\epsilon-ia)^{\alpha}}\right\}
\end{equation}
since $a>0$ being non-zero we can further evaluate (\ref{evintyields}) by putting $\epsilon=0$
to arrive at
\begin{equation}
\label{evintyieldsfinale}
{\cal J}_{\alpha}(a)=\int_{0}^{a}b_{\alpha}(x){\rm d}x=\frac{(\alpha-1)!}{\pi a^{\alpha}}\sin{(\frac{\pi \alpha}{2})}
\end{equation}
which compensates indeed (\ref{intainf}) so that ${\cal J}_{\alpha}(a)+{\cal I}_{\alpha}(a)=0$ thus (\ref{bexplaalphint}) indeed is fulfilled with the remarkable property

\begin{equation}
\label{evlimits}
\lim_{a\rightarrow 0+}{\cal J}_{\alpha}(a)=\lim_{a\rightarrow 0+}\int_{0}^{a}b_{\alpha}(x){\rm d}x=sign(\sin{(\frac{\pi \alpha}{2})})\times\infty \,,\hspace{2cm} \alpha >0
\end{equation}
is diverging for $\alpha>0$ as $a^{-\alpha}$!
That means if we imagine for a moment $|b_{\alpha}|$ being a mass density, that the total mass
concentrated in the interval $0 \leq x \leq a$ is the same as the total mass in the interval
$a<x<\infty$. We conclude that $b_{\alpha}(x)$ has, in dependence on $\alpha$, several oscillatoric peaks of alternate signs localized for $\epsilon\rightarrow 0+$ infinitely close to $x=0$. The integral over these localized oscillations up to a finite limit $a>0$ compensates integral
(\ref{intainf}). The strength of these oscillatoric peaks are ``infinitely stronger'' as that one of a $\delta$-function. The closed form expression of the $b_{\alpha}$ which takes into account this singularity is therefore (\ref{intalpha})
which fulfils for instance the "desired" property of a fractional derivative of order $\alpha >0$ (e.g. \cite{jumarie}) that it yields zero when applied to a constant.
This property requires that

\begin{equation}
\label{probfrac}
\int_0^{a}f(x){\rm d}x= Const \times a^{-\alpha}
\end{equation}
which is obviously not solved for $\alpha > 0$ by any real valued power-function $f(x)=const\, x^{-\alpha -1} \in \R$ since its primitive in (\ref{probfrac}) diverges at $x=0$. In contrast $b_{\alpha}(x)$ solves (\ref{probfrac})
\begin{equation}
\label{verfy}
Re\left\{\int_{0}^{\infty}\frac{{\rm d}x}{(\epsilon -ix)^{\alpha+1}}\right\}= Re\left\{\frac{(-i)}{\alpha}\frac{1}{(\epsilon -ix)^{\alpha}}|_0^{\infty}\right\}=0 \,,\hspace{2cm}\alpha >0
\end{equation}
for any $\epsilon >0$ since the lower limit of the integral is purely imaginary and the upper limit vanishing for $\alpha >0$. Hence (\ref{intalpha}) defines for $\alpha >0$ an integration kernel for a fractional derivative which yields zero when applied to a constant which is due to the vanishing of (\ref{bexplaalphint}) which we can indeed verify directly.
We analyze in appendix \ref{fractional} further properties of $b_{\alpha}$.

Let us now return to the functions $q_n(x)$ of (\ref{qnencore}): Taking into account (\ref{intalpha}) have with $\alpha=n\delta >-1$ and with

\begin{equation}
\label{qndelta}
b_{n\delta}(x)=\frac{{\cal L}^{n}}{A_{\delta}^n}\delta(x)= \frac{(n\delta)!}{\pi}
\lim_{\epsilon \rightarrow 0+}Re\left\{\frac{i^{n\delta+1}}{(|x|+i\epsilon)^{n\delta+1}} \right\} \,,\hspace{2cm} n\delta >-1
\end{equation}
where $A_{\delta}$ is given in (\ref{Adelta2}).
For $n=1$ we obtain

\begin{equation}
\label{q1deltaa}
b_{\delta}(x)=\frac{1}{A_{\delta}}{\cal L}\delta(x)=\frac{\delta !}{\pi}\lim_{\epsilon \rightarrow 0+} Re\left\{\frac{i^{\delta+1}}{(|x|+i\epsilon)^{\delta+1}} \right\}
\end{equation}
being in accordance with $q_1(x)={\cal L}\delta(x)=-\Delta_{(\delta,h,\zeta)}\delta(x)$
obtained by evaluation of (\ref{laplace}) or (\ref{laplaceb}) for $x\neq 0$.
From (\ref{qndelta})
one can recover the {\it static} Green's function by (\ref{staticgreen})
\begin{equation}
\label{qndeltagrenn}
q_{-1}(x)=\frac{b_{-\delta}(x)}{A_{\delta}}={\cal L}^{-1}\delta(x)=\frac{(-\delta)!}{\pi A_{\delta}}Re\left\{\frac{i^{1-\delta}}{(|x|+i\epsilon)^{1-\delta}} \right\} \,,\hspace{2cm} 0<\delta<2
\end{equation}
which indeed recovers expression (\ref{reform3}) for this Green's function $q_{-1}(x)=g(x)$. (\ref{qndeltagrenn}) takes into account in a "correct" way the singular behavior at $x=0$ and is therefore more general as (\ref{reform3}) holding only for $x\neq 0$. Whereas the $b_{n\delta}(x)$ hold for $n\delta >-1$, the functions $q_n(x)=A_ {\delta}^nb_{n\delta}(x)$ are only physically meaningful where $A_{\delta} >0$ and where (\ref{laplace}) exists namely $0<\delta<2$. The range of validity of the $b_{\alpha}(x)$ can be {\it mathematically} further extended to all $\alpha < -1$ if one utilizes instead of
(\ref{gammafu}) the regularized definition of the $\Gamma$-function (\ref{expresgen}).

The performed analysis gives reason to the following observation: There is a {\it general} link between non-integer $\alpha$ including odd integers $\alpha$ and the non-locality and self-similarity of physical properties. This behavior is also reflected by the non-locality of our Laplacian
(\ref{laplace}). Intuitively in a wider sense one is tempted to state: {\it even} integer-exponents $\alpha$ (case {\bf (iii)}) correspond to local (``regular''), non-integer and odd exponents $\alpha$ (case {\bf (iv)}) to non-local, self-similar (``irregular'')\footnote{The meaning of the term ``irregular'' is here used in the sense of ``erratic'', it does {\it not} mean the absence of rules.} physical behavior.

\section{Conclusions}

We have analyzed an elastic medium with self-similar elastic energy-density. The source of the physics in this continuum is the self-similar {\it Laplacian} (\ref{laplace}) having the form of a non-local convolution with a power-law convolution kernel. This Laplacian can be represented in terms of fractional derivatives and is obtained by the continuous limiting case of the discrete self-similar Laplacian introduced in \cite{michel}. The resulting equations of motion are self-adjoint partial {\it fractional} differential equations which are only in the range $0<\delta<2$ physically meaningful.

We deduced closed form solutions {\bf (i)} for the {\it static} Green's function (displacement field due to a unit
$\delta$-force), {\bf (ii)} for the Cauchy problem (two convolution kernels "propagators"), {\bf (iii)} for the solution of a self-similar diffusion problem. The solution was obtained in terms of a convolution kernel solving the initial value problem.
It has been found that the obtained convolution kernel represents normalized L\'evi-stable probablity distributions with inifinite variances (mean fluctuations) in their range of definition $0<\delta<2$.

As a spin-off result we obtained a set of functions $b_{n\delta}(x)$ which we denoted as {\it self-similar potentials} appearing as regularized Fourier transforms of $|k|^{\alpha}$. These functions are to be conceived as {\it generalized functions}
in the distributional sense of Gel'fand and Shilov \cite{gelfand}. Self-similar potentials are special solutions of the self-similar Poisson's equation and play in statics the analogous role as Newtonian potentials for the traditional Poisson equation.

It has been demonstrated that the concept of self-similar functions introduced in \cite{michel} is useful to describe material systems with self-similar interactions. Our approach opens the door to a continuum description of a rich fond of new physical especially dynamic phenomena in such systems. Especially interesting are the propability-distributions obtained as solutions of the self-similar (fractional) diffusion equation (\ref{diffussioneq1}). These probability-distributions are L\'evi-stable which are known to be a characteristic feature of the statistics of irregular and erratic motions characterized by discontinuous trajectories (L\'evi-flights) \cite{mandel,mandel2}. It is mentionworthy that our diffussion model recovers the range $0<\delta<2$ of admissible L\'evi parameter $\delta$ given previously in the literature (e.g. \cite{mandel,mandel2}).

We have further formulated the 1D continuity equation for
the self-similar diffusion problem and formulated the self-similar probability-flux (Eq. (\ref{laplacebflux})).
An interesting open point which can be treated by the present approach remains the formulation of the non-local self-similar Gauss-theorem in dimensions higher than one, and more general to formulate continuity equations and related problems in self-similar continua embedded into the 2D or 3D space.

The present approach can be a point of departure to answer such questions in a rather rigorous way. For example the present approach can be considered as a point of departure to any general self-similar field theory such as continuum mechanics and electrodynamics. We hope the present paper stimulates work in such directions.

\section{Appendix}
\subsection{Fractional derivative}
\label{fractional}

In view of the analysis of the last section we can easily recall a {\it fractional derivative} $D_x^{\alpha}f(x)$
which fulfills for $\alpha > 0$ all desirable properties of a derivative. We call positive orders (fractional) {\it derivatives} and negative orders (fractional) {\it integrals}. Let us consider the case $\alpha >-1$ which can however be extended to $\alpha < -1$ in the spirit of generalized functions \cite{gelfand} as outlined in appendix \ref{balphamin1}.

Let us first of all determine
\begin{equation}
\label{frac1diffeq2}
y_{\alpha}(x)=D_x^{\alpha}\delta(x) \,,\hspace{2cm} \alpha > -1
\end{equation}
where we call $y_{\alpha}(x)$ a fractional {\it derivative}. For $\alpha=0$ the solution of should be $y_{\alpha=0}(x)=\delta(x)$.

From (\ref{frac1diffeq2}) follows then for the definition of
this fractional derivative applied on a function $f(x)$ the convolution

\begin{equation}
\label{deffracint}
D_x^{\alpha}f(x)=\int_{-\infty}^{\infty}y_{\alpha}(x-\tau)f(\tau){\rm d}\tau \,,\hspace{2cm} \alpha >-1
\end{equation}
with the kernel $y_{\alpha}(x)$ (\ref{Dxalpha3}) to be determined.
Further we assume that $D_x^{\alpha}e^{ikx}=(ik)^{\alpha}e^{ikx}$ so we can write $y_{\alpha}(x)$
write as a Fourier integral

\begin{equation}
\label{Dxalpha}
y_{\alpha}(x)=\frac{1}{2\pi}\int_{-\infty}^{\infty}e^{ikx}(ik)^{\alpha}{\rm d}k
\end{equation}
which we can rewrite

\begin{equation}
\label{Dxalpha2}
y_{\alpha}(x)=\frac{1}{2\pi}\lim_{\epsilon\rightarrow 0+}\int_{0}^{\infty}e^{-\epsilon k}k^{\alpha}\left(e^{i(kx+\frac{\alpha\pi}{2})}+e^{-i(kx+\frac{\alpha\pi}{2})}\right){\rm d}k
\end{equation}
which can be further written as

\begin{equation}
\label{Dxalpha3}
y_{\alpha}(x)=\frac{\alpha !}{\pi}\lim_{\epsilon\rightarrow 0+}Re\left\{\frac{i^{2\alpha+1}}{(x+i\epsilon)^{\alpha+1}} \right\}   \,,\hspace{2cm} \alpha > -1
\end{equation}

For $x\neq 0$ (\ref{Dxalpha3}) assumes the form

\begin{equation}
\label{ybalpha}
y_{\alpha}(x)= -\frac{\alpha !}{\pi|x|^{\alpha +1}}\sin{\pi\alpha}
\end{equation}

We verify that for integer $\alpha=0,1,2,..\in \N_0$ (\ref{ybalpha}) is vanishing for $x\neq 0$ and
(\ref{Dxalpha3}) yields

\begin{equation}
\label{endres2balpha}
y_n(x)=\frac{(-1)^n n !}{\pi}\lim_{\epsilon\rightarrow 0+} Re\left\{\frac{i}{(x+i\epsilon)^{n+1}} \right\} =
\frac{d^n}{dx^n}\frac{1}{\pi} \lim_{\epsilon\rightarrow 0+} Re\left\{\frac{i}{(x+i\epsilon)} \right\}
=\frac{d^n}{dx^n}\delta(x)
\end{equation}
reproducing in the integer-case the correct localized the integer-order derivatives.

\subsection{Convergence of series (\ref{Qxt1}), (\ref{Qxtt2})}
\label{convergence}
It still remains to verify the convergence of the series (\ref{Qxt1}) and (\ref{Qxtt2}):
To this end we consider series (\ref{Qxtt2}) which has the form

\begin{equation}
\label{verfyconv}
\frac{\partial }{\partial t}Q(x,t)=\delta(x)-\frac{1}{\pi|x|}Im \left\{P\left(\frac{A_{\delta}t^2e^{\frac{\pi i \delta}{2}}}{|x|^{\delta}}\right)\right\}
\end{equation}
where $q_0(x)=\delta(x)$ and $P$ is only a function of $\xi=\frac{A_{\delta}t^2e^{\frac{\pi i \delta}{2}}}{|x|^{\delta}})$ where $P$ is given by

\begin{equation}
\label{Pfu}
P(\xi)=\sum_{n=1}^{\infty}\frac{(n\delta)!}{(2n)!}\xi^n\,,\hspace{2cm} 0<\delta <2
\end{equation}
In order to judge the convergence of this series the asymptotic representation of $\frac{(n\delta)!}{(2n)!}$ for $n>>1$ will be useful. To this end we put $s(n)=2n(1-\frac{\delta}{2}))$ with $n\delta=2n-s$

\begin{equation}
\label{assympto}
\frac{(n\delta)!}{(2n)!}= \frac{(2n-s(n))!}{2n!} \rightarrow \frac{1}{(2n)^{2n(1-\frac{\delta}{2})}}\,,\hspace{2cm} 0 < \delta < 2
\end{equation}

From (\ref{assympto}) we can determine the radius of convergence $\rho_c$ of the series
(\ref{Pfu}) where the series $\sum_{n_0}^{\infty} a_n\xi^n$ converges absolutely for $|\xi| < \rho_c$ with

\begin{equation}
\label{radcon}
\rho_c=\lim_{n\rightarrow\infty}|\frac{a_n}{a_{n+1}}|=\lim_{n\rightarrow\infty}(2ne)^{1-\frac{\delta}{2}} =\infty \,,\hspace{1cm} 0< \delta < 2
\end{equation}
since $1-\frac{\delta}{2}>0$. Hence series (\ref{Pfu}) converges for all $\xi$ absolutely which guarantees
the convergence of (\ref{Qxt1}) and (\ref{Qxtt2}) for all $x$ and $t$ for $\delta$ being in the interval $0< \delta < 2$.

\subsection{Convergence of series (\ref{propa3})}
\label{conv2}

We consider the convergence of (\ref{propa3})
\begin{equation}
\label{propa3b}
W(x,t)= \sum_{n=0}^{\infty}(-1)^n\frac{t^n}{n!}\,q_n(x) \,,\hspace{2cm} 0<\delta<1
\end{equation}
which we can rewrite in the form
\begin{equation}
\label{propa4}
W(x,t)= \delta(x)-\frac{1}{\pi|x|}Im\{w(\frac{A_{\delta}te^{\frac{\pi i\delta}{2}}}{|x|^{\delta}})\}
\end{equation}

where $w(\xi)$ with $\xi=\frac{A_{\delta}t e^{\frac{{\pi i \delta}}{2}}}{|x|^{\delta}}$ has the representation

\begin{equation}
\label{wreprep1}
w(\xi)=\sum_{n=1}^{\infty}(-1)^n\frac{(n\delta)!}{n!}\xi^n \,,\hspace{2cm} 0<\delta<1
\end{equation}

The radius of convergence $r_c$ of this series is

\begin{equation}
\label{radcon2}
r_c=\lim_{n\rightarrow\infty}\frac{(n\delta)!}{n!}\frac{(n+1)!}{((n+1)\delta)!}=(n+1)\frac{(n\delta)!}{((n+1)\delta)!}\rightarrow n^{1-\delta}\delta^{-\delta}\rightarrow\infty \,,\hspace{1cm} 0<\delta< 1
\end{equation}

Hence (\ref{wreprep1}) converges for all $\xi$ absolutely uniformly. Hence (\ref{propa3}) converges in the entire domain $-\infty < x < \infty$ and $t>0$ for $\delta$ being in the interval $0<\delta< 1$.

\subsection{Some remarks on the Fourier-integral (\ref{intalph}) for $\alpha < -1$}
\label{balphamin1}

A further observation is important in  $b_{\alpha}(x)$ of (\ref{intalph}) which is the Fourier transform of $|k|^{\alpha}$ exists for $\alpha> -1$. For exponents $\alpha < -1$ the integral (\ref{intalph}) diverges. However, we can define a Fourier
transform of $Re\{(-i)^{\alpha}(|k|+i\epsilon)^{\alpha}\}$ in the limiting case $\epsilon\rightarrow 0+$ which exists also for $\alpha < -1$.

We generalize the definition of $b_{\alpha}(x)$ (\ref{intalph}) for $\alpha < -1$ by replacing
$|k|^{\alpha} \rightarrow \frac{1}{\cos{\frac{\pi\alpha}{2}}}Re(\epsilon-i|k|)^{\alpha})$ in the form
\begin{equation}
 \label{balpahalph}
b_{\alpha}(x)=\lim_{\epsilon\rightarrow 0+} \frac{1}{\pi\cos{\frac{\pi\alpha}{2}}}Re\left\{(-i)^{\alpha}\int_0^{\infty}(k+i\epsilon)^{\alpha}\cos{kx}\,{\rm d}k\right\} \,,\hspace{2cm} \alpha \in \R \notin \pm 1, \pm 3,..\pm (2n+1),..
\end{equation}
which exists $\alpha \in \R$ where the zeros of the cosine have to be excluded. For $\alpha > -1$ (\ref{balpahalph}) is identical with (\ref{intalph}) and for $\alpha < -1$ it complements the representation
(\ref{intalph}) which diverges in this case.
Let us especially evaluate (\ref{balpahalph}) in the ``forbidden'' case $\alpha < -1$: Taking into account

\begin{equation}
 \label{putting}
(\epsilon -i|k|)^{\alpha}=\frac{1}{\Gamma(-\alpha)}\int_0^{\infty}e^{-\tau(\epsilon -i|k|)}\tau^{-\alpha-1}{\rm d\tau}
\end{equation}
existing for $\alpha < -1$ we obtain the representation

\begin{equation}
 \label{represent1}
b_{\alpha}(x)=\lim_{\epsilon\rightarrow 0+} \frac{1}{\pi\cos{\frac{\pi\alpha}{2}}\Gamma(-\alpha)}Re\left\{
\int_0^{\infty}{\rm d}\tau\, e^{-\epsilon\tau}\tau^{-\alpha-1}\int_0^{\infty}\cos{kx}\, e^{ik\tau}{\rm d}k\right\}
\end{equation}
which we evaluate by using 
\begin{equation}
 \label{usingrel}
Re \int_0^{\infty}\cos{kx}\, e^{ik\tau}{\rm d}k = \frac{\pi}{2}\delta(\tau-|x|)
\end{equation}
and arrive at

\begin{equation}
 \label{intermsalph}
b_{\alpha}(x)=\frac{1}{2\cos{\frac{\pi\alpha}{2}}}\frac{|x|^{-\alpha-1}}{\Gamma(-\alpha)} \,,\hspace{2cm} \alpha < -1 , \notin -1,  -3,..-(2n+1),..
\end{equation}
where $\Gamma(-\alpha)$ is for $\alpha < -1$ well defined in (\ref{gammafu}). 

By taking into account 
the Euler relation (\ref{expresgen}) we arrive at

\begin{equation}
\label{endres}
b_{\alpha}(x)=-\frac{\alpha !}{\pi|x|^{\alpha+1}}\sin{\frac{\alpha\pi}{2}} \,,\hspace{2cm} \alpha < -1 , \notin -1,  -3,..-(2n+1),..
\end{equation}
which has the same form as expression (\ref{bexplaalph}) which holds for $\alpha > -1$ but with the generalized definition
of $\alpha !$ given in (\ref{expresgen}). In conclusion (\ref{endres}) holds for $\alpha \in \R \notin \pm 1, \pm 3,..\pm (2n+1),..$.
For cases $\alpha$ being even integers, expression (\ref{endres}) is not valid, we refer in this context further to \cite{gelfand} where the Fourier transforms $1/|k|^{2n+1}$ ($n=0,1,2..$) are further analyzed.

\end{document}